\documentclass[10pt]{article}  
\usepackage{color}
\usepackage{dcolumn}   
\usepackage{bm}        
\usepackage{amssymb}   
\usepackage{amsmath,textcomp}
\usepackage{xfrac}
\usepackage[utf8]{inputenc}
\usepackage{float}
\usepackage{authblk}
\usepackage{hyperref}
\usepackage[margin = 1in]{geometry}
\hypersetup{%
    pdfborder = {0 0 0}
} 

\begin{document}


\title{Quantum back action cancellation in the audio band}

\author[1]{Jonathan Cripe}
\author[1]{Torrey Cullen}
\author[2]{Yanbei Chen}
\author[3]{Paula Heu}
\author[3]{David Follman}
\author[3,4]{Garrett D. Cole}
\author[1,*]{Thomas Corbitt}
\affil[1]{Department of Physics \& Astronomy, Louisiana State University, Baton Rouge, LA 70803}
\affil[2]{Theoretical Astrophysics 350-17, California Institute of Technology, Pasadena, California 91125}
\affil[3]{Crystalline Mirror Solutions LLC and GmbH, Santa Barbara, CA, and Vienna, Austria}
\affil[4]{Vienna Center for Quantum Science and Technology (VCQ), Faculty of Physics, University of Vienna, A-1090 Vienna, Austria}
\affil[*]{email:tcorbitt@phys.lsu.edu}

	\date{\today}
\date{\today}

\maketitle

\begin{abstract}

We report on the cancellation of quantum back action noise in an optomechanical cavity. We perform two measurements of the displacement of the microresonator, one in reflection of the cavity, and one in transmission of the cavity. We show that measuring the amplitude quadrature of the light in transmission of the optomechanical cavity allows us to cancel the back action noise between 1 kHz and 50 kHz, and obtain a more sensitive measurement of the microresonator's position. 
To confirm that the back action is eliminated, we measure the noise in the transmission signal as a function of circulating power.
By splitting the transmitted light onto two photodetectors and cross correlating the two signals, we remove the contributon from shot noise and measure a quantum noise free thermal noise spectrum.
Eliminating the effects of back action in this frequency regime is an important demonstration of a technique that could be used to mitigate the effects of back action in interferometric gravitational wave detectors such as Advanced LIGO. 
\end{abstract}



\section{Introduction}

Over the past century, interferometers have been used to perform increasingly sensitive measurements for a wide variety of applications. From the pioneering work of Michelson and Morley's attempts to measure the aether \cite{Michelson} to the recent discoveries of gravitational waves by the worldwide gravitational wave detector network \cite{GW150914, BNS_detection, O1O2_catalog}, interferometers have been used to probe the minuscule displacements of mechanical objects. Shot noise, a consequence of the particle nature of light, has been a fundamental limitation to the sensitivity of gravitational wave interferometers. As the power used in advanced gravitational wave detectors is increased in order to reduce the impact of shot noise, the detectors are approaching the regime in which quantum back action (QBA) begins to limit the sensitivity. QBA arises from the fluctuations of the ponderomotive force imparted by the light used in the measurement \cite{Caves_1980, Braginsky_book, Caves1981}. It has only been recently, however, that experiments have been able to observe the subtle quantum effects of QBA \cite{Purdy, Teufel, Purdy_room_T, Sudhir}.

With experiments in the QBA-limited regime, ideas for how to manipulate and ultimately remove the effects of QBA become experimentally accessible. Several recent experiments have successfully evaded QBA using two mechanical oscillators \cite{Korppi_BAE, BA_evasion_Moller}. Another proposed method of removing QBA is a variational measurement, in which the readout quadrature is chosen such that the measurement is free of QBA \cite{Kimble, Variation}. The variational readout technique utilizes correlations between quadratures of the light to effectively cancel the QBA, resulting in a measurement limited only by shot noise \cite{Regal_variational}. Ordinarily, this cancellation must be done in a frequency-dependent way using an optical filter cavity. In this work, we take advantage of the correlations created in a detuned optomechanical cavity with a strong optical spring to cancel the effects of QBA in a single, frequency independent quadrature, specifically the amplitude quadrature of the light that is transmitted through the cavity.

In a previous experiment, the displacement was measured of a low-loss, single-crystal microresonator \cite{cole08, cole12, cole13, cole14, Singh_PRL, Cripe_QRPN} that forms one mirror of an optomechanical Fabry-Perot cavity, and the result showed that QBA was the dominant source of its motion between 10 kHz and 50 kHz \cite{Cripe_QRPN}. In this work, we build upon the previous results and show that we are able to perform a QBA-free measurement of the displacement of the microresonator by detecting the amplitude quadrature of the light that is transmitted through the optomechanical cavity.
We present two displacement measurements of the microresonator, one in reflection of the cavity and one in transmission, and show that the measurement done in transmission is free of QBA. The transmission measurement shows a measurable reduction in displacement noise between the frequencies of 2 kHz and 50 kHz compared to the measurement done in reflection.


\section{Theory}

To understand how we are able to perform a QBA-free measurement of the position of the microresonator, it is useful to consider the equation of motion of the movable mirror, given by
\begin{equation}
m \frac{d^2x}{dt^2} = -\frac{2\delta P}{c}+F_{\mathrm{ext}} = -\frac{2}{c}\left[\left(\frac{\partial P}{\partial a_1}\right)_x a_1+\left(\frac{\partial P}{\partial x}\right)_{a_1}x\right]+F_{\mathrm{ext}},
\end{equation}
where $m$ and $x$ are the mass and position of the microresonator, $\delta P$ is the power fluctuation, $F_{\mathrm{ext}}$ is an external force, and $c$ is the speed of light.
The first term is the fluctuating radiation pressure acting on the mirror, driven by incoming vacuum fluctuations in the amplitude quadrature, $a_1$. The second term arises in a detuned cavity and creates the optical spring effect \cite{17}. We may then write, in the frequency domain
\begin{equation}
-m\Omega^2 x = -\frac{2}{c}\left(\frac{\partial P}{\partial a_1}\right)_x a_1 - m \Omega_{\mathrm{OS}}^2 x+F_{\mathrm{ext}},
\label{EOM_freq}
\end{equation}
where $\Omega_{\mathrm{OS}}$ is the optical spring frequency, and we have used the relation $\left(\frac{\partial P}{\partial x}\right)_{a_1}=m\Omega_{OS}^2$.
Solving for the position, $x$, and power fluctuations, $\delta P$, leads to
\begin{eqnarray}
x &=& \frac{1}{m(\Omega^2 -  \Omega_{\mathrm{OS}}^2)}\frac{2}{c} \left(\frac{\partial P}{\partial a_1}\right)_x a_1-\frac{1}{m(\Omega^2 -  \Omega_{\mathrm{OS}}^2)}F_{\mathrm{ext}},
\end{eqnarray}
and
\begin{eqnarray}
\delta P &=& \frac{m c x \Omega^2}{2}+ \frac{c}{2} F_{\mathrm{ext}}=\frac{\Omega^2}{\Omega^2-\Omega_{\mathrm{OS}}^2}\left(\frac{\partial P}{\partial a_1}\right)_x a_1-\frac{c}{2}\frac{\Omega^2}{\Omega^2 -  \Omega_{\mathrm{OS}}^2}F_{\mathrm{ext}}+ \frac{c}{2} F_{\mathrm{ext}}\\
&=&\frac{\Omega^2}{\Omega^2-\Omega_{\mathrm{OS}}^2}\left(\frac{\partial P}{\partial a_1}\right)_x a_1-\frac{c}{2}\frac{\Omega_{OS}^2}{\Omega^2 -  \Omega_{\mathrm{OS}}^2}F_{\mathrm{ext}}.
\label{deltaP}
\end{eqnarray}
As evident from the first term in Equation \ref{deltaP}, for frequencies far below the optical spring resonance, such that $\Omega \ll \Omega_{\mathrm{OS}}$, the power fluctuations in the cavity are reduced by the factor $\left(\Omega_{\mathrm{OS}} / \Omega\right)^2$. Measuring the amplitude quadrature of the light exiting the cavity in transmission is equivalent to measuring the same quadrature as the intracavity field because the transmitted carrier has the same phase as the intracavity carrier. Since the intracavity amplitude fluctuations are highly suppressed, the amplitude fluctuations in the transmitted field will also be strongly suppressed. Thus, the vacuum fluctuations that promptly reflect from the end mirror dominate, and the total noise will be close to shot noise. Therefore, a signature of QBA is not observed in this measurement. We do note, however, that external (classical) forces do couple to the intracavity power and to the amplitude of the transmitted field. Thus, measuring the amplitude of the transmission does provide a measure of the mirror displacement, but without a component from QBA.

\begin{figure}
\center
\includegraphics[width= \columnwidth]{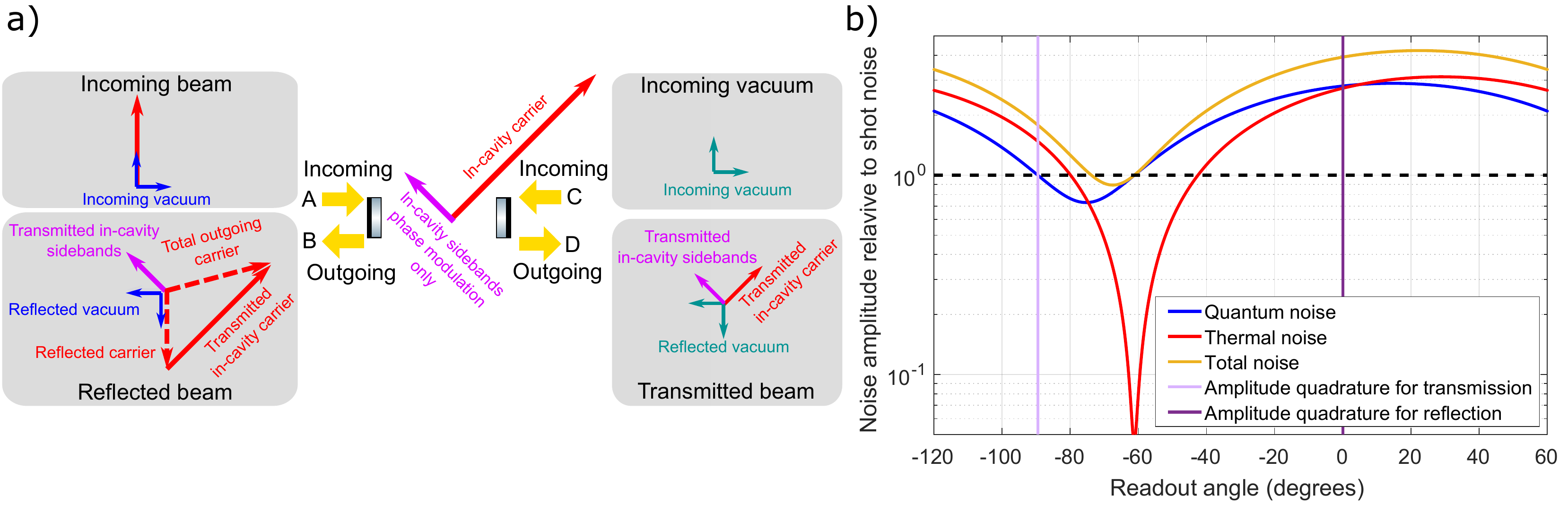}
\caption{a) Diagram explaining the relationship between the quadratures of the intracavity light, transmitted light, and reflected light. Input $A$ contains the incoming vacuum and the carrier of the laser field, while input $C$ only consists of vacuum fluctuations. The transmitted beam, $D$, is composed of the reflected vacuum fluctuations from $C$ and a small amount of the intracavity field, whose amplitude fluctuations are highly suppressed as discussed in the text. The reflected light, $B$, is made up of the promptly reflected carrier and vacuum fluctuations as well as the outgoing intracavity light. The total reflected carrier is the vector sum of the promptly reflected and transmitted intracavity beams.  
b) Modeled quantum noise, thermal noise, and the total noise relative to shot noise (dotted horizontal line) at 20 kHz as a function of readout angle with a cavity circulating power of 155 mW and detuning of -0.55 linewidths. The amplitude quadrature for light detected in reflection of the cavity is set at 0 degrees and marked with a vertical line. The amplitude quadrature for the light detected in transmission of the microresonator is marked with the vertical line at $-89$ degrees where the quantum noise curve is equal to shot noise.
Note that the figure does not include technical noise sources and is meant to serve more as a pedagogical tool that the total noise can be reduced rather than a detailed noise budget, which is included later in the paper.}
\label{quad_readout}
\end{figure}

If the cavity is measured in reflection, however, the situation is very different. The intracavity field exiting at the input coupler must mix with the field that promptly reflects from the input coupler. These two fields are at a different phase that depends on the cavity detuning. Measuring the amplitude quadrature of the reflected field will therefore measure a different quadrature than doing so in transmission. Figure \ref{quad_readout}a illustrates the relationship between the intravacity light and the cavity inputs and outputs.

To provide an additional view of how to perform a measurement without QBA, we show in Figure \ref{quad_readout}b how quantum noise (which includes both shot noise and QBA), thermal noise, and the total noise scales with respect to shot noise as a function of the readout angle in which the light is detected. In the numerical model, using the two photon formalism \cite{Corbitt_mathematical}, we use values of 155 mW for the circulating power and 0.55 linewidths for the detuning to match the experimental values described below.
As shown in Figure \ref{quad_readout}b, there are two readout angles at which the quantum noise curve is equal to one. At these points, the quantum noise is equal to the shot noise and contains no contribution from QBA.  One of these quadratures corresponds to a quadrature in which the measurement is insensitive to the position of the microresonator. This is seen at a readout angle of about $-60^{\circ}$ where there is a large dip in the contribution from thermal noise and the total noise curve is approximately equal to shot noise. 
The second quadrature that has the quantum noise equal to shot noise is at approximately $-90^{\circ}$ and is marked by a vertical line in Figure \ref{quad_readout}b. This quadrature is sensitive to the position of the microresonator, and the total noise is equal to the quadrature sum of thermal noise and shot noise and does not contain a contribution from QBA. Performing a measurement at this quadrature is equivalent to detecting the amplitude quadrature of the light in transmission of the microresonator as outlined in the theoretical calculations above.
As a note, changing these values for the circulating power and detuning in the model will change the angle between the amplitude quadrature for the reflected and transmitted light.


\section{Experiment and Results}

In contrast to our previous measurement of QBA in which we detected the light that was reflected by the optomechanical cavity \cite{Cripe_QRPN}, we instead measure the light that is transmitted through the cavity. This change in the detection scheme allows us to measure the quadrature that does not contain QBA, as described in the section above. 

\begin{figure}
\center
\includegraphics[width= \columnwidth]{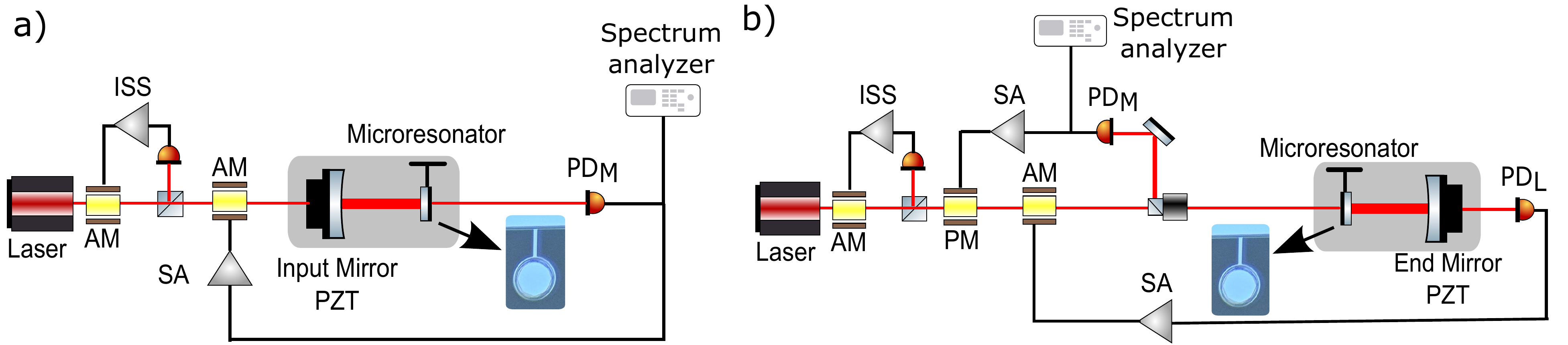}
\caption{Experimental setup for the a) back action cancellation measurement in transmission of the cavity and b) benchmark measurement in reflection of the cavity \cite{Cripe_QRPN}. a) Light from a 1064 nm Nd:YAG laser is passed through two amplitude modulators (AM) before being injected into the optomechanical cavity, which sits on a suspended optical breadboard to reduce seismic motion and is housed in a vacuum chamber at $10^{-7}$ Torr (shown in shaded gray).
A micrograph of the single-crystal microresonator, comprising a 70-$\mu$m diameter GaAs/AlGaAs mirror pad supported by a GaAs cantilever, is included in the diagram.
An intensity stabilization servo (ISS) is used to stabilize the laser power to shot noise by feeding back to the first AM. The light transmitted through the optomechanical cavity is detected by $\mathrm{PD_M}$. The signal from $\mathrm{PD_M}$ is sent through a servo amplifier (SA) before being sent to the second AM to lock the cavity. The signal from $\mathrm{PD_M}$ is also sent to a spectrum analyzer for further analysis.}     
\label{Setup_BAE}
\end{figure}

A comparison of the experimental setups for the measurement in transmission and reflection is shown in Figure \ref{Setup_BAE}.
In both setups, light is injected into the optomechanical cavity after passing through an intensity stabilization servo loop. 
The optomechanical cavity is slightly less than 1 cm long and is formed by a 1 cm radius of curvature macroscopic mirror with a transmission of 50 ppm and a high-reflectivity single-crystal microresonator with a transmission of 250 ppm. The microresonator consists of a roughly 70 $\mu$m diameter mirror pad suspended from a single-crystal GaAs cantilever with a thickness of 220 nm, width of 8 $\mu$m, and length of 55 $\mu$m. The mirror pad is made up of 23 pairs of quarter-wave optical thickness GaAs/Al$_{ 0.92}$Ga$_{0.08}$As layers for a transmission of $250$ ppm and exhibits both low optical losses and a high mechanical quality factor \cite{cole08, cole12, cole13, cole14, Singh_PRL}. The microresonator has a mass of 50 ng, a natural mechanical frequency of 
$876$ Hz,
and a measured mechanical quality factor of $16,000$ at room temperature (295 K). The cavity has a finesse of $13,000$ and linewidth (HWHM) of 580 kHz \cite{Cripe_QRPN}. 

The difference between the measurement in transmission and reflection, is that in the QBA cancellation (transmission) measurement, the light is injected through the macroscopic input mirror and is detected in transmission of the microresonator by the photodetector labeled $PD_\mathrm{M}$. In the reflection measurement, the laser is injected through the microresonator and is detected in reflection at $PD_\mathrm{M}$. The difference between injecting the light through the macroscopic mirror and  microresonator means that the cavity is under-coupled for the transmission measurement in Figure \ref{Setup_BAE}a and over-coupled for the reflection measurement in Figure \ref{Setup_BAE}b.
An additional difference between the two setups is that we are able to simplify the control scheme for locking the cavity in transmission and only require a single feedback loop to an amplitude modulator \cite{Cripe_RPL} rather than a second feedback loop that requires a phase modulator.

We perform two displacement measurements of the microresonator, one in transmission of the cavity as shown in Figure \ref{Setup_BAE}a, and one in reflection of the cavity, as shown in Figure \ref{Setup_BAE}b. 
We first perform a measurement of the displacement noise of the cavity with the light injected on the microresonator side of the cavity and detected in reflection. This measurement provides us with a measurement of the displacement noise when the contribution from QBA is included. The cavity orientation and feedback for this measurement is shown in Figure \ref{Setup_BAE}b and mimics that of the QBA measurement in \cite{Cripe_QRPN}. To determine the optical parameters including the detuning, circulating power, and intracavity loss for the measurement, we have found that the most accurate method is using measurements of the optical spring \cite{Cripe_QRPN}.
We measured the optical spring frequency to be 142 kHz in this orientation. The reflection measurement was performed with an input power of 50 $\mu$W, and a detuning of 0.55 $\pm 0.05$ linewidths. The measurement of the optical spring, along with the cavity detuning and input power, was used to calculate the circulating power of 155 mW $\pm 10$ mW inside the cavity and the intracavity loss to be 200 ppm $\pm 10$ ppm.

After performing the reflection measurement, we switch which side of the cavity the light is incident on and inject the light from the macroscopic mirror side of the cavity. Detecting the amplitude quadrature of the light in transmission of the cavity with $PD_\mathrm{M}$ allows us to have a QBA-free measurement of the position of the microresonator. 
Between the two measurements, we carefully keep the circulating power and detuning as close to constant as experimentally possible so that we can make a fair comparison between the results of the two measurements and look for a reduction in the measured displacement noise.
We performed the transmission measurement with the same optical spring frequency as the reflection measurement. We set the detuning to 0.50 $\pm 0.05$ linewidths by finding the maximum optical spring frequency for the given input power of 220 $\mu$W and then use the optical spring measurement to calibrate the optical parameters of circulating power of 155 mW $\pm 10$ mW, and intracavity loss 180 ppm $\pm 10$ ppm. 



\begin{figure}
\center
\includegraphics[width= \columnwidth]{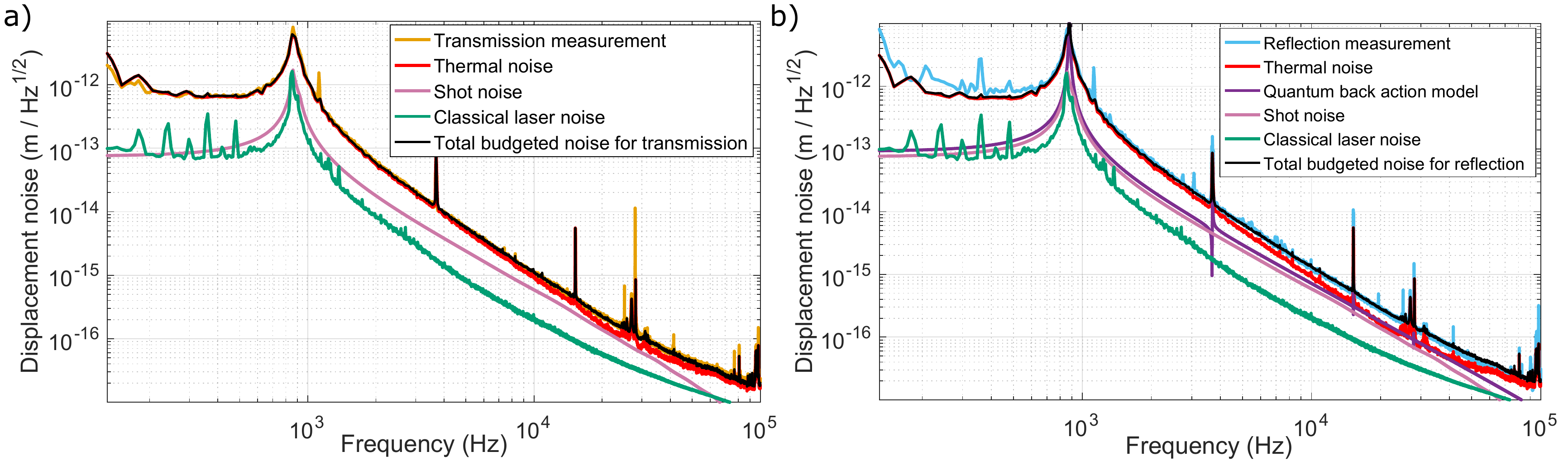}
\caption{Measured and budgeted displacement noise spectra taken in transmission (a) and reflection (b) of the cavity. The individual noise sources for each measurement are added in quadrature to calculate the total budgeted noise. The coupling of the higher order mechanical modes of the microresonator are reduced with respect to their magnitudes in \cite{Cripe_QRPN} as a result in improvements in our ability to align the optical beam onto the nodal points of those modes.}      
\label{Noise_budgets}
\end{figure}

The results from both the reflection measurement and the back action cancellation transmission measurement are shown in Figure \ref{Noise_budgets}. 
Figure \ref{Noise_budgets}a is a detailed noise budget of the noise sources that contribute to the transmission measurement, and Figure \ref{Noise_budgets}b is for the reflection measurement. The total budgeted noise for each measurement is calculated by adding in quadrature all of the individual noise sources that contribute to the total noise.

In the transmission measurement, shown in Figure \ref{Noise_budgets}a, the largest noise source is thermal noise, which is plotted in red, with the sum of the measured shot noise and dark noise having the next largest magnitude.
Contributions from the measured classical laser intensity noise and classical laser frequency noise \cite{Willke} are added in quadrature and included in the curve labeled classical laser noise. The measurement of thermal noise shown in Figure \ref{Noise_budgets} was done with a circulating power of 10 mW inside the cavity so as to minimize the magnitude of the shot noise in the measurement \cite{Cripe_QRPN}. When each of the individual noise sources are added in quadrature, the sum of the individual noise sources agrees with the measured displacement noise across the entire measurement band from 100 Hz to 100 kHz. 


We also calculate the total budgeted noise for the reflection measurement and plot the result in Figure \ref{Noise_budgets}b. The noise budget for the reflection measurement contains the same noise sources as those in the transmission measurement, but it also includes a contribution from the QBA that is present in the reflection measurement. The inclusion of the QBA in the noise budget for the reflection measurement accounts for the increased noise level as compared to the transmission measurement which does not contain QBA.

\begin{figure}
\center
\includegraphics[width=.95\columnwidth]{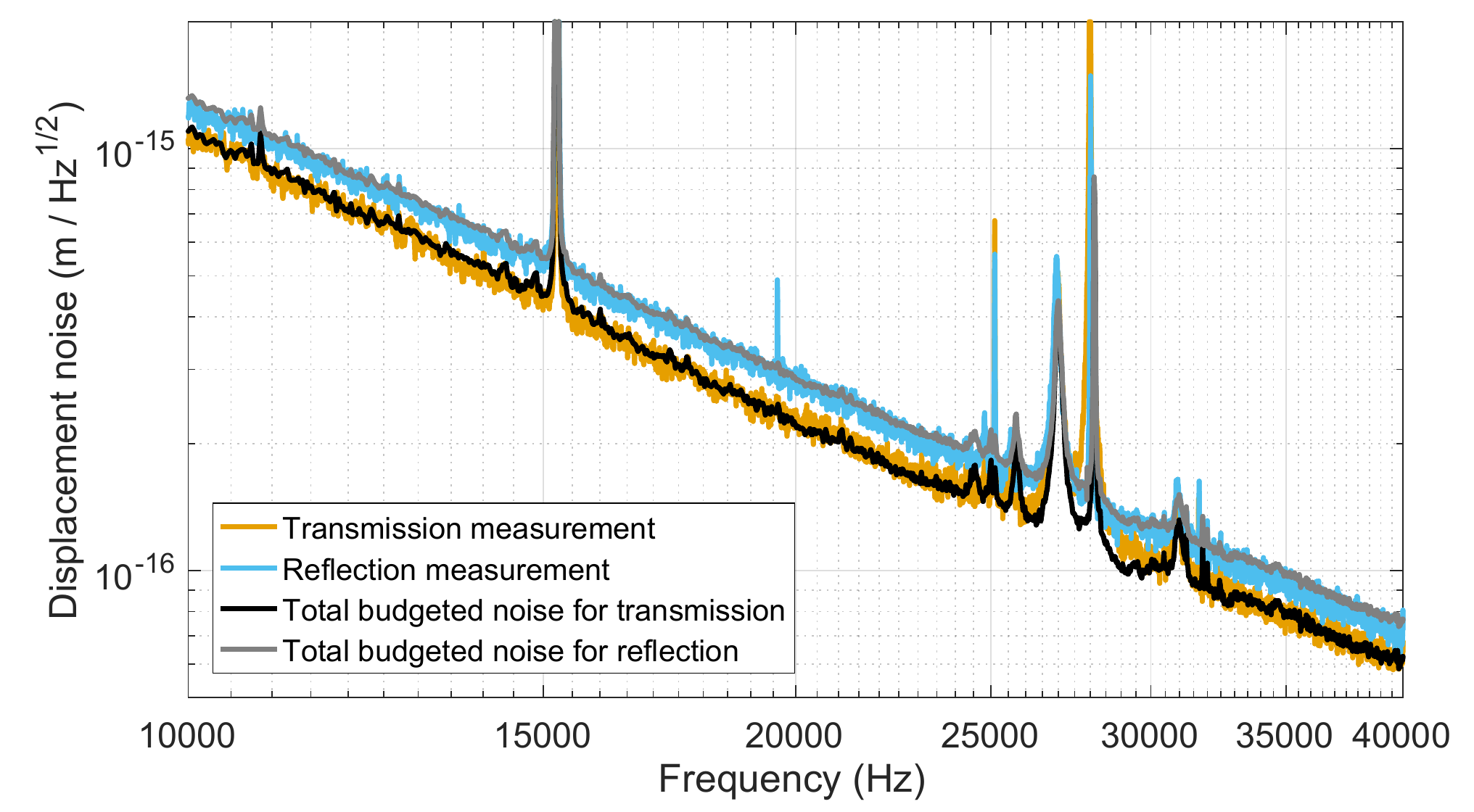}
\caption{Measured and budgeted total noise for the transmission and reflection measurements. The data shown here is the same as that plotted in Figure \ref{Noise_budgets} but plotted over a smaller frequency range to make the reduction in noise more visible.}     
\label{Zoom}
\end{figure}

To see the difference between the two noise measurements more clearly, we include a magnified version of the noise budgets in Figure \ref{Zoom}, where we focus on the region between 10 kHz and 40 kHz and plot the total measured and budgeted noise for each measurement. Comparing the measured displacement noise for the transmission and reflection measurements in Figures \ref{Noise_budgets} and \ref{Zoom} shows that the measured noise for the transmission measurement is reduced between approximately 1 kHz and 50 kHz, with the maximum reduction of 20 \% or about 2 dB at 20 kHz as compared to the reflection measurement.

To confirm that we are canceling the QBA, we modify the transmission measurement by splitting the transmitted light across two photodetectors, and cross correlating their outputs to remove the effects of shot noise. The results of this measurement are shown in Figure \ref{Thermal}, and details of the cross correlation measurement are included in the Supplementary Material section. Performing the cross correlation and removing the effects of shot noise allows us to compare measurements taken with different circulating power levels and demonstrate that the measured noise remains constant and at the level of thermal noise. If QBA were present in the measurements, we would expect to see an increase in the measured noise that scales with the square root of power \cite{Caves_1980}.

In addition to confirming that the QBA is canceled in the measurements, measuring the cross correlation also allows us to perform a quantum noise free measurement of the thermal noise of the microresonator. This measurement also serves as a validation that the temperature of the microresonator does not significantly increase between low and high circulating power measurements. Physical heating of the device would lead to an increase in thermal noise, which scales with the square root of temperature. The results shown in Figure \ref{Thermal} reveal that there is not an increase in thermal noise for measurements spanning a factor of 10 in circulating power.

\begin{figure}
\center
\includegraphics[width= 0.8 \columnwidth]{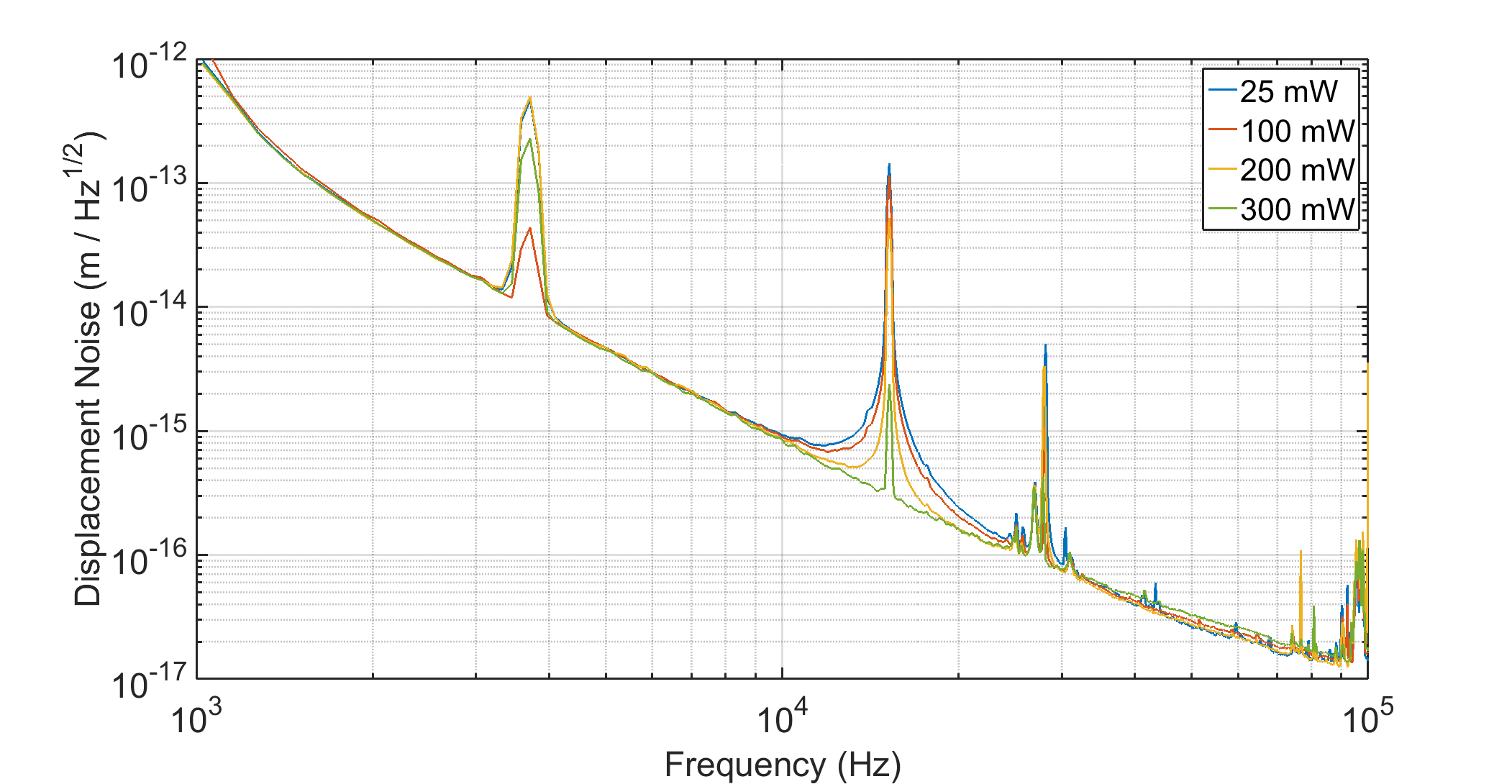}
\caption{Quantum noise free thermal noise spectrum measured at 4 different power levels. The measurement is performed by splitting the transmitted light onto two photodetectors and cross correlating their outputs (see Supplementary Material). The measured thermal noise remains constant across the range of powers at frequencies away from the resonances of the higher order modes. At frequencies near the higher order modes, the measured noise changes as a result of the change in cavity alignment due to the deflection of the cantilever by radiation pressure \cite{Cripe_QRPN}. Realigning the beam at high power recovers the same thermal noise as measured with low power. There is no signature of QBA as expected.}     
\label{Thermal}
\end{figure}

\section{Conclusion}

We have performed two displacement noise measurements of a microresonator in an optomechanical cavity and have shown that the measured noise is reduced when the light is detected in transmission of the cavity as compared to in reflection of the cavity. The reduction in noise is a result of the cancellation of QBA in the spectrum of the transmitted light. The measured noise is reduced between 2 kHz and 50 kHz by up to 2 dB, and is consistent with the theoretical calculations and the measurements of the individual noise sources. 
To confirm the QBA cancellation, we perform a cross correlation with two photodetectors in transmission of the cavity and show that the measured noise remains constant for measurements spanning a factor of 12 in power. Our results demonstrate the cancellation of QBA in the audio band at frequencies relevant to current and future interferometric gravitational wave detectors.

This work was supported by National Science Foundation grants PHY-1150531 and PHY-1806634. 
The microresonator manufacturing was carried out at the UCSB Nanofabrication Facility.
This document has been assigned the LIGO document number {LIGO-P1800380}. 


\section{Supplementary Material}

Here we outline the cross correlation measurement used to perform the quantum noise free thermal noise measurement. First, to eliminate quantum radiation pressure noise, we measure the intensity of the transmitted light, which is insensitive to QBA. Then, we split the light transmitted through the cavity onto two photodetectors using a beamsplitter. By taking advantage of the fact that thermal noise is correlated between the two detectors, whereas shot noise is not, we can measure only the thermal noise by looking at the cross power spectral density between the two photocurrents. The optical field in the amplitude quadrature for the first and second PD may be written
\begin{eqnarray}
a_1 = v_1 + \alpha x_{th}\\
b_1 = v_2 + \alpha x_{th},
\end{eqnarray}
where $v_1$ and $v_2$ correspond to uncorrelated vacuum fluctuations, and $\alpha$ is the coupling constant for thermal moiton.
A cross power spectral density taken between the two outputs of the PDs gives
\begin{equation}
|<S_{12}>| = |\alpha|^2 S_{th},
\label{Eq_CC}
\end{equation}
where $S_{th}$ is the thermal noise spectrum. Notably, all the terms involving vacuum fluctuations go to $0$ because they are uncorrelated with each other, and with the thermal noise. In principle, this allows for the true thermal noise spectrum to be obtained, with no contribution from quantum noise.  

In our experiment, however, we require the transmitted signal to be fed back to an AM for locking the cavity, which can create additional correlations. In this case, the feedback to the cavity reinjects the noise on only one of the PDs back onto the amplitude quadrature of the input light. In this case, the fields for each PD may be written as
\begin{eqnarray}
a_1 = v_1 + \alpha x_{th} - \frac{G}{1+G}\left(v_1 + \alpha x_{th}\right) = \frac{v_1 + \alpha x_{th}}{1+G}\\
b_1 = v_2 + \alpha x_{th}-\frac{G}{1+G}\left(v_1 + \alpha x_{th}\right)=v_2 + \frac{\alpha x_{th}}{1+G}-\frac{G}{1+G}v_1,
\end{eqnarray}
where $G$ is the open loop gain for the feedback loop \cite{Cripe_RPL}. 
Then the cross power spectral density between the two photocurrents is
\begin{equation}
|<S_{12}>| = \frac{|\alpha|^2 S_{th}}{1+G} + \frac{G}{\left(1+G\right)^2},
\end{equation}
where we have normalized the PSD to shot noise. Therefore, there is an extra noise contribution equal to shot noise multiplied by the factor $G / (1+G)$. For this extra noise to be significant, shot noise must be similar to or greater in magnitude than thermal noise, which only occurs at the high power levels. In addition, the loop gain must satisfy $|G|\gtrapprox1$. For the frequencies of interest here, these requirements are not simultaneously satisfied, and the extra noise is small.


\bibliographystyle{ieeetr}

\end{document}